\documentclass[12pt]{iopart}

\usepackage{color}
\usepackage{amssymb}
\usepackage{graphicx}
\usepackage{mathrsfs}
\usepackage{CJK}
\usepackage[figuresright]{rotating}
\usepackage{epstopdf}
\usepackage[colorlinks=black,
            linkcolor=black,
            anchorcolor=black,
            citecolor=black]{hyperref}

\usepackage{url}

\newtheorem{exam}{Example}[section]
\newtheorem{df}{Definition}[section]

\newtheorem{thm}{Theorem}[section]
\newtheorem{cor}{Corollary}[section]
\newtheorem{lem}{Lemma}[section]
\newtheorem{rmk}{Remark}[section]

\begin{document}

%\title[Outer synchronization of two-layer networks with delayed nodes and
%noise perturbation]{Outer synchronization of two-layer networks with delayed nodes and
%noise perturbation}
\title[Counterpart synchronization of duplex networks]{Counterpart synchronization of duplex networks with delayed nodes and
noise perturbation}
%\author{Content \& Services Team}
%
%\address{IOP Publishing, Temple Circus, Temple Way, Bristol BS1 6HG, UK}

\author{Xiang Wei$^{1,2}$ \& Xiaoqun~Wu$^{1,*}$ \&Jun-an Lu$^1$ \& Junchan Zhao$^{3}$}

\address{$^1$School of Mathematics and Statistics, Wuhan University, Wuhan 430072, China\\
$^2$Department of Engineering, Honghe University, Honghe 661199, China\\
$^3$School of Mathematics and Statistics, Hunan University of Commerce, Changsha 410205,  China}
\ead{xqwu@whu.edu.cn(Xiaoqun~Wu)}
\vspace{10pt}
\begin{indented}
\item[]July 2015
\end{indented}

%\begin{abstract}
%This document describes the  preparation of an article using \LaTeXe\ and
%\verb"iopart.cls" (the IOP Publishing \LaTeXe\ preprint class file).
%This class file is designed to help
%authors produce preprints in a form suitable for submission to any of the
%journals listed in table~\ref{jlab1} on the next page.  You are not obliged to use this class file---we accept
%submissions using all common \LaTeX\ class and style files.  The \verb"iopart.cls"
%class file is supplied merely as a convenience for those authors who find it useful.
%This document gives both general advice that applies whatever class file you use, and specific advice
%that applies if you choose to use \verb"iopart.cls".
%
%We also accept submissions in Word format.  Instructions for Word submissions are available via the `Author Guidelines' link at \verb"http://authors.iop.org".
%
%If you have any queries about this document or any aspect of preparing your article for submission please contact us at the e-mail address given above.
%\end{abstract}
\begin{abstract}
  In the real world, many complex systems are represented not by single networks but rather by sets of interdependent ones. In these specific networks, nodes in one network mutually interact  with nodes in other networks.  This paper focuses on a simple representative case of two-layer networks (the so-called duplex networks) with unidirectional inter-layer couplings. That is,  each node in one network   depends on a counterpart  in the other network.  Accordingly, the former network is called the response layer and the latter network is the drive layer.  Specifically,   synchronization between each node in the drive layer and its counterpart in the response layer ~(counterpart synchronization, or CS) of this sort of duplex networks with delayed nodes
 and noise perturbation is investigated.    Based on the LaSalle-type invariance principle, a control technique is proposed  and a sufficient condition is developed for realizing counterpart synchronization of duplex networks.Furthermore,  two corollaries are derived as special cases. In addition, node dynamics within each layer can be various and topologies of  the two layers  are not necessarily identical.
Therefore, the proposed synchronization method can be applied to a wide  range of multiplex  networks. Numerical examples are provided to illustrate the feasibility and effectiveness of the results.\\
 {\bf{Keywords:}}{~complex network, duplex,  counterpart synchronization, stochastic perturbation}
\end{abstract}
%\keywords{magnetic moment, solar neutrinos, astrophysics}

%\maketitle

\section{Introduction}\label{Intro}
Complex networks abound in almost every aspect of science and technology.  Examples include  the Internet, the World Wide Web, social networks, metabolic networks, food webs, networks of citations between papers, among many others~ \cite{Kurths2003,Boccaletti2006,Osipov2009}. Synchronization is one of the most common phenomena in nature that interacting  nodes can reach a coherent state, and it has been extensively investigated and discussed in the past two
decades~\cite{Watts1998,Nishikawa2003,Motter2005,Pecora1998,Huang2009,Wang2002a,Wang2002}. For example, Pecora et al.~\cite{Pecora1998} used the
master stability function (MSF) approach to analyze the stability
of the synchronous state in coupled systems, Huang et al.~\cite{Huang2009} classified  synchronization
 into five categories based on the MSF approach, Wang and Chen  investigated synchronization in small-world networks~\cite{Wang2002a} and scale-free~\cite{Wang2002} networks.

Many literatures, including the above-mentioned ones, are primarily focused on synchronization within  single networks that do not interact with other networks. However,  many real-world networks often interact with and depend on each other.
 For example, people in a society interact with each other via their family relationship, friendship,  or  formal work-related acquaintanceship~\cite{D2014Networks}. Countries in the global economic system also interact via various international relations. Transportation  depends  on   air traffic networks,   railway networks and   road traffic networks. Obviously, in describing and dealing with such problems, the multiplex network representation would be more appropriate than the single network.  Not surprisingly,  multiplex networks have attracted enormous  attention in the past few years in various fields of application.  For example,  Xiong et al.~\cite{Xiong2014Correlation} analysed the correlation between the information diffusion process and the opinion evolution process and found obvious interaction between the two processes.  Liu et al.~\cite{Liu2014Modeling} investigated preferred degree networks and their interactions, and found dramatically  different behaviors between two very similar networks.

Counterpart synchronization describes the individuals in one network behave coherently with their   counterparts in  other associated  networks, so it represents
harmonious coexistence of nodes in multiplex networks.  This sort of synchronization in a duplex network can also be taken as the so-called  outer synchronization between two networks and has attracted wide attention. For example,  Wu et al. investigated  generalized outer synchronization between two different complex dynamical networks by employing  nonlinear control~\cite{Wu2009}. In particular, this   synchronization has been widely applied in topology identification of complex networks. To name just a few, Wu~\cite{Wu2008} and Zhao et al.~\cite{Zhao2010} employed complete outer synchronization to identify  topologies for weighted complex networks, Zhang  et al. \cite{Zhang2014} et al. adopted generalized outer synchronization to recover network structures.

Meanwhile, time delays are unavoidable in complex networks due to finite information processing and propagation speeds. They extensively exist in the real world, such as communication networks, gene regulatory networks, and electrical power grids.
Time delays greatly influence behaviors of dynamical systems. Many literatures are focused on synchronization and control of complex networks with coupling delay among different nodes~\cite{Zheng2012, Wang2010}.

Noise is another important factor affecting  behaviors of dynamical systems, as it is inevitable due to environmental disturbance and uncertainties. Generally, noise is harmful. However, the presence of noise sometimes plays a positive role~\cite{Wang2010a}, such as in inducing synchronization~\cite{Nagai2010} and in facilitating topology identification of complex networks~\cite{Wu2011, Wu2012a}.

Motivated by  above discussions, we investigate CS of duplex networks with delayed nodes and noise perturbation.  Based on the LaSalle-type invariance principle for stochastic differential delay equations, we design adaptive controllers to synchronize nodes of the response  layer to their counterparts in the drive layer, and put forward some sufficient conditions for guaranteeing  CS.

The rest of this paper is organized as follows. Modeling of  duplex networks and some preliminaries are introduced in Section~\ref{pre}. Sufficient conditions for CS in duplex networks are presented in Section~\ref{main}. In Section~\ref{simulations}, two numerical examples are provided to illustrate the feasibility and effectiveness of our method. Finally, some conclusions are drawn in Section~\ref{con}.

\textbf{Notation:} Some necessary notations used throughout the paper are introduced. $\mathbf{x}^\top$ (or $\mathbf{A}^\top$) denotes the transpose of a vector $\mathbf{x}$ (or a matrix $\mathbf{A}$), $\parallel\mathbf{x}\parallel_2$ is the Euclidean-norm of  $\mathbf{x}$, $\otimes$ represents the Kronecker product, $\mathbb{R}^n$ is the $n$-dimensional real space, $I_n\in \mathbb{R}^{n\times n}$ represents an identity matrix of order $n$, $C^n[a,b]~(a, b\in\mathbb{R},~a<b)$ represents the $n-$order continuously differentiable function space in $[a,b]$.

\section{ Modeling and preliminaries}\label{pre}
\begin{figure}[!hbt]
 % \begin{tabular}{c}
  \centering
  \mbox{\includegraphics*[width=0.75\columnwidth,clip=0]{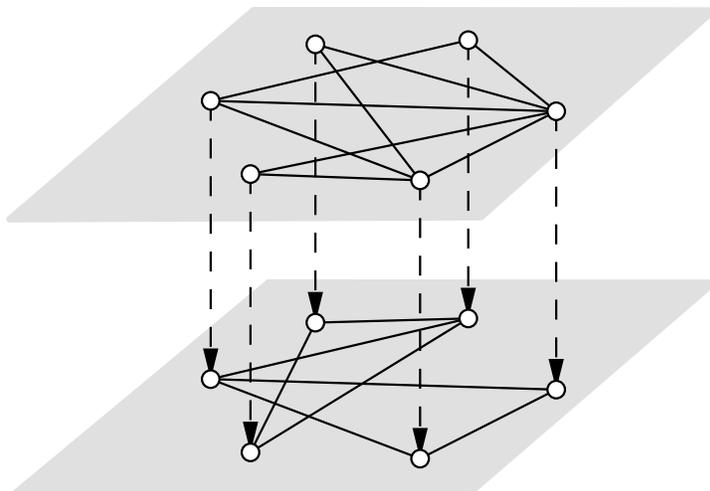}}
   %\mbox{\includegraphics*[width=0.45\columnwidth,clip=0]{11.eps}}
%\end{tabular}
{\caption{A duplex network with unidirectional inter-layer couplings. Each node in the upper layer is unidirectionally connected to a counterpart  in the lower layer.  The topologies of the two layers can be different, representing the individual sort of intra-layer interactions. }\label{topology2}}
  %\includegraphics[width=4.2cm]{11.jpg}\hspace*{3ex}
 % \includegraphics[width=8.4cm]{12.jpg}
%  \caption{Total synchronization error between networks (\ref{new-driven}) and (\ref{new-response}).}\label{E0}
%  \centering
%%\begin{center}
%\includegraphics[width=8.0cm ]{13.jpg}\hspace*{3ex}
%\includegraphics[width=8.0cm ]{14.jpg}
%\includegraphics[ width=6cm]{ouhe_aver.eps}
\end{figure}

Consider a duplex network  consisting of $\mathbf{N}$ nodes in each layer, as shown in Fig. \ref{topology2}.   For convenience,  take the upper layer  as the  drive layer and the lower layer which is dependent on signals from the drive layer as the response layer.
We are concerned about the impact of delayed nodes and noise caused
by control input. Thus a drive layer consisting of $N$ linearly coupled nodes  is described by
%\begin{eqnarray} \label{old-drive}
%%d\mathbf{x}_i(t)=[\mathbf{f}_i(\mathbf{x}_i(t))+\sum_{j=1}^{N}c_{ij}\Gamma \mathbf{x}_j(t)]dt,\quad i=1,2,...,N.
%d\mathbf{x}_i(t)=[\mathbf{f}_i(t,\mathbf{x}_i(t),\mathbf{x}_i(t-\tau(t)))+\sum_{j=1}^{N}c_{ij}\Gamma \mathbf{x}_j(t)]dt,~i=1,2,...,N,
%\end{eqnarray}
\begin{eqnarray}\label{new-driven}
%\begin{aligned}
%\aligned
d\mathbf{x}_i(t)=[\mathbf{f}_i(t,\mathbf{x}_i(t),\mathbf{x}_i(t-\tau(t)))+\sum_{j=1}^{N}c_{ij}\Gamma \mathbf{x}_j(t)]dt,~i=1,2,...,N,
%\end{aligned}
%\aligned
\end{eqnarray}
and the response layer with   control input is  given by
%\begin{equation}\label{old-response}
%%d\mathbf{y}_i(t)=[\mathbf{f}_i(\mathbf{y}_i(t))+\sum_{j=1}^{N}d_{ij}\Gamma \mathbf{y}_j(t)+\mathbf{u}_i(t)]dt,\quad i=1,2,...,N.
%\end{equation}
\begin{eqnarray}\label{new-response}
%\begin{aligned}
\fl
d\mathbf{y}_i(t)=[\mathbf{f}_i(t,\mathbf{y}_i(t),\mathbf{y}_i(t-\tau(t)))
+\sum_{j=1}^{N}d_{ij}\Gamma \mathbf{y}_j(t)+\mathbf{u}_i(t)]dt \nonumber\\
+\mathbf{\sigma}_i(t,\mathbf{e}_i(t),\mathbf{e}_i(t-\tau(t)))d\mathbf{w}(t),~i=1,2,...,N.
%\end{aligned}
\end{eqnarray}
Here, $\mathbf{x}_i(t)=(\mathbf{x}_{i1},...,\mathbf{x}_{in})^\top\in \mathbb{R}^n$ and $\mathbf{y}_i(t)=(\mathbf{y}_{i1},...,\mathbf{y}_{in})^\top\in \mathbb{R}^n$ are state vectors, $\mathbf{u}_i(t)$ is the control input for  node $i$, $\mathbf{f}_i:\mathbb{R}_+\times\mathbb{R}^n \times \mathbb{R}^n\rightarrow \mathbb{R}^n$ is a continuously differentiable function   determining the dynamical behavior of   node $i$,   $\Gamma=(a_{ij})_{n\times n}\in \mathbb{R}^{n\times n}$ is the inner coupling matrix, and
$\mathbf{C}=(c_{ij})_{N\times N}\in \mathbb{R}^{N\times N}$ is the  coupling
configuration matrix representing the coupling strength and the topological structure of network (\ref{new-driven}), with $c_{ij}$ being defined as follows: if there is a link from node $j$ to node
$i~(i\neq j)$,    $c_{ij}\neq0$; otherwise, $c_{ij}=0$. The diagonal elements of matrix $\mathbf{C}$ is $c_{ii}=-\sum_{j=1,j\neq i}^{N}c_{ij}$ for $i=1,2,..., N$. $\mathbf{D}=(d_{ij})_{N\times N}\in \mathbb{R}^{N\times N}$ is the coupling
configuration matrix of network (\ref{new-response}), which has the same meaning as that of $\mathbf{C}$.
%In this paper, we are concerned with the effect of delayed nodes and noise caused by control input. Therefore, the following two-layer networks are considered, network (\ref{new-driven}) as drive layer and network (\ref{new-response}) as response layer.
%\begin{eqnarray}\label{new-driven}
%%\begin{aligned}
%%\aligned
%d\mathbf{x}_i(t)=[\mathbf{f}_i(t,\mathbf{x}_i(t),\mathbf{x}_i(t-\tau(t)))+\sum_{j=1}^{N}c_{ij}\Gamma \mathbf{x}_j(t)]dt,~i=1,2,...,N,
%%\end{aligned}
%%\aligned
%\end{eqnarray}
%\begin{eqnarray}\label{new-response}
%%\begin{aligned}
%d\mathbf{y}_i(t)=[\mathbf{f}_i(t,\mathbf{y}_i(t),\mathbf{y}_i(t-\tau(t)))
%+\sum_{j=1}^{N}d_{ij}\Gamma \mathbf{y}_j(t)+\mathbf{u}_i(t)]dt \nonumber\\
%+\mathbf{\sigma}_i(t,\mathbf{e}_i(t),\mathbf{e}_i(t-\tau(t)))d\mathbf{w}(t),~i=1,2,...,N,
%%\end{aligned}
%\end{eqnarray}
 $\tau(t)$ denotes  time delay of nodes,   $\mathbf{e}_i(t)=\mathbf{y}_i(t)-\mathbf{x}_i(t)$.
The noise term in network (\ref{new-response}) is   utilized to describe perturbation caused by the control input process  influenced by environmental fluctuations \cite{Sun2013}. In particular, $\sigma_i:\mathbb{R}^+\times \mathbb{R}^n\times \mathbb{R}^n \rightarrow \mathbb{R}^{n\times m}$ is called
the noise intensity matrix, $\mathbf{w}(t)=(w_1(t),...,w_m(t))^\top$ is an $m$-dimensional Brownian motion defined on a complete
probability space $(\Omega,\mathcal{F},P)$ with a natural filtration $\{\mathcal{F}_{t} \}_{t\geq0}$.
%Accordingly, $\mathbf{\dot w}(t)$ is $m$-dimensional white noise.

Throughout this paper, we make the following assumptions:\\
%\begin{hypo}
(H1)\label{H1}
The noise intensity function $\mathbf{\sigma}_i(t,\mathbf{x},\mathbf{y})~(i=1,2,...,N)$ satisfies the Lipschitz condition and there exists positive constants $p,q$ such that
\begin{eqnarray} \label{H1}
trace(\mathbf{\sigma}_i^\top\mathbf{\sigma}_i)\leq p\mathbf{x}^\top \mathbf{x}+q\mathbf{y}^\top \mathbf{y}.
\end{eqnarray}
Moreover $\sigma(t,0,0)\equiv 0$.\\
%\end{hypo}
%\begin{hypo}
(H2)\label{H2}
There exists a positive constant $M$ such that
\begin{eqnarray}
%\begin{aligned}
\fl
 \|\mathbf{f}_i(t,\mathbf{x}_i(t),\mathbf{x}_i(t-\tau(t)))-\mathbf{f}_i(t,\mathbf{y}_i(t),\mathbf{y}_i(t-\tau(t)))  \| \nonumber\\
\leq{M}[\|(\mathbf{x}_i(t)-\mathbf{y}_i(t))\|^2+\|(\mathbf{x}_i(t-\tau(t))-\mathbf{y}_i(t-\tau(t)))\|^2]^\frac {1}{2}.
%\end{aligned}
\end{eqnarray}
(H3)\label{H3}
\emph{$\tau(t)$ is a differentiable function with }
  \begin{eqnarray}
  0\leq \dot \tau(t)\leq \mu<1.
  \end{eqnarray}
  \emph{Obviously, this assumption is
ensured if the delay $\tau(t)$ is constant.}
%\end{hypo}

Our purpose is to design proper controllers so that the noise-perturbed response layer (\ref{new-response}) can reach CS with the drive layer (\ref{new-driven}). For this purpose,   some necessary concepts and a lemma  of stochastic
differential equations are   presented.

 Consider the following $n$-dimensional stochastic differential delay equation:
\begin{eqnarray}
\label{formula8}
d\mathbf{z}(t)=\phi(t,\mathbf{z}(t),\mathbf{z}(t-\tau))dt+\varphi(t,\mathbf{z}(t),\mathbf{z}(t-\tau))d\mathbf{w}
\end{eqnarray}
on $t\geq 0$ with an initial value $\xi \in C_{\mathcal{F}_0}^\mu([-\tau,0],\mathbb{R}^n)$, where $C_{\mathcal{F}_0}^\mu([-\tau,0],\mathbb{R}^n)$  represents the family of all ${\mathcal{F}_0-}$measurable bounded $C([-\tau,0],\mathbb{R}^n)-$valued random variables,  the measurable functions $\phi,\varphi:[0,+\infty]\times \mathbb{R}^n \times \mathbb{R}^n\rightarrow \mathbb{R}^n$ satisfy the locally Lipschitz condition and the linear growth condition. It is  known that Eq.(\ref{formula8}) has a unique solution for any initial value $\xi$ that is denoted by $\mathbf{z}(t,\mathbf{\xi})$ on $t\geq -\tau$.

Let $C^{1,2}(\mathbb{R}_+\times \mathbb{R}^n,\mathbb{R}_+)$ denote the family of all non-negative functions $V(t,\mathbf{z})$ on $\mathbb{R}_+\times \mathbb{R}^n$, which are continuously once differentiable in $t$ and twice differentiable in $\mathbf{z}$. For each $V\in C^{1,2}(\mathbb{R}_+\times \mathbb{R}^n,\mathbb{R}_+)$, the   diffusion operator $\mathcal{L}V$ associated to (\ref{formula8}) acting on  $C^{1,2}(\mathbb{R}_+\times \mathbb{R}^n,\mathbb{R}_+)$ is defined by
\begin{eqnarray} \label{L-op}
\mathcal{L}V=\frac{\partial V}{\partial t}+\frac{\partial V}{\partial z}\cdot \phi+\frac{1}{2}trace[\varphi^T\frac{\partial^2 V}{\partial^2 z}\cdot \varphi],
\end{eqnarray}
where $\partial V/\partial \mathbf{z}=(\partial V/\partial z_1,...,\partial V/\partial z_n),\partial^2 V/\partial^2 \mathbf{z}=(\partial^2 V/\partial z_i\partial z_j)_{n\times n}$.

\begin{lem}\label{lem1}
(A Lasalle-type invariance theorem for stochastic differential equations \cite{Mao1999}). Assume that both $\phi(t,\mathbf{u},\mathbf{v})$ and $\varphi(t,\mathbf{u},\mathbf{v})$ are locally bounded in $(\mathbf{u},\mathbf{v})$ while uniformly bounded in t. Assume also that there are functions $V\in C^{1,2}(\mathbb{R}_+\times \mathbb{R}^n,\mathbb{R}_+)$, $\gamma \in L^1(\mathbb{R}_+,\mathbb{R}_+)$, and $\omega_1,\omega_2\in \mathbf{C}(\mathbb{R}^n,\mathbb{R}_+)$ such that

$$\mathcal{L}V(t,\mathbf{u},\mathbf{v})\leq \gamma(t)-\omega_1(\mathbf{u})+\omega_2(\mathbf{v}),\quad \forall(t,\mathbf{u},\mathbf{v})\in \mathbb{R}_+\times \mathbb{R}^n\times \mathbb{R}^n,$$  $$\omega_1(\mathbf{u})\geq \omega_2(\mathbf{u}), \quad \forall \mathbf{u}\in \mathbb{R}^n$$
and
$$\lim\limits_{\parallel \mathbf{u}\parallel \to +\infty}\inf\limits_{0\leq t\leq \infty}{V(t,\mathbf{u})=\infty}.$$

%\lim\limits_{t\rightarrow \infty}\|x_{j}(t)-x_{i}(t)\|=0
  % $\lim\limits_{t\rightarrow \infty}\|v_{j}(t)-v_{i}(t)\|=0,\forall~i,j=1,2\cdots ,N.\notag$

Then $Ker(\omega_1-\omega_2)\neq \emptyset$ and for every initial value $\xi \in \mathbf{C}_{\mathcal{F}_0}^\mu([-\tau,0],\mathbb{R}^n)$,  the solution $\mathbf{z}(t,\xi)$ of Eq. (\ref{formula8}) has the following property:
 $$\lim\limits_{t \to \infty }{dist\{\mathbf{z}(t;\xi),Ker(\omega_1-\omega_2) \}}=\textbf{0}~~a.s.$$
Moreover, if $Ker(\omega_1-\omega_2)=0$, then for every $\xi \in C_{\mathcal{F}_0}^\mu([-\tau,0],\mathbb{R}^n)$, $\lim\limits_{t \to \infty }{\mathbf{z}(t;\xi)=0}$ a.s.
\end{lem}

\section{Sufficient conditions for CS of duplex networks}\label{main}
In this section, we will first give the definition on top of duplex networks.
\begin{df}
 The duplex network formed by the drive layer  (\ref{new-driven}) and the response layer (\ref{new-response}) is said to almost surely achieve CS   if
 \begin{eqnarray}\label{error}
 \mathbf{e}_i(t)=\lim\limits_{t \to \infty }\|\mathbf{y}_i(t)-\mathbf{x}_i(t)\|=0,\quad i=1,2,...,N.
 \end{eqnarray}
\end{df}

With the network models and the definition given previously, we arrive at the following main theorem.
\begin{thm} \label{thm}
 Let  (H1), (H2) and (H3) hold. The response layer (\ref{new-response}) can almost surely achieve CS with the drive layer (\ref{new-driven}) with the following control scheme:
 \begin{eqnarray} \label{controller}
  \mathbf{u}_i(t)=\sum_{j=1}^{N}b_{ij}(t)\Gamma \mathbf{y}_j(t)-\mathbf{g}_i(t)\mathbf{e}_i(t), i=1,2,...,N,
\end{eqnarray}
\begin{eqnarray} \label{adaptive}
\dot g_i(t)=k_i\|\mathbf{e}_i(t)\|^2, \dot b_{ij}(t)=-\mathbf{e}_i^\top(t)\Gamma \mathbf{y}_j(t),i=1,2,...,N,
\end{eqnarray}
where $k_i>0~(i=1,2,...,N)$ are arbitrary constants, $b_{ij}(t), g_i(t)~(i,j=1,2,...,N)$ are adaptive parameters updating with network dynamics.
\end{thm}

\textbf{Proof.} Since $\mathbf{e}_i(t)=\mathbf{y}_i(t)-\mathbf{x}_i(t)$,   dynamics of the synchronization error between counterparts in  layers (\ref{new-driven}) and (\ref{new-response}) can be written as follows:
\begin{eqnarray}\label{d-error}
%\begin{aligned}
\fl
d\mathbf{e}_i(t)=[\mathbf{f}_i(t,\mathbf{y}_i(t),\mathbf{y}_i(t-\tau(t)))-\mathbf{f}_i(t,\mathbf{x}_i(t),\mathbf{x}_i(t-\tau(t)))+\sum_{j=1}^N(d_{ij}\Gamma \mathbf{y}_j(t)-c_{ij}\Gamma \mathbf{x}_j(t))\nonumber\\
\fl+\mathbf{u}_i(t)]dt 
+\sigma_i(t,\mathbf{e}_i(t),\mathbf{e}_i(t-\tau(t)))d\mathbf{w}(t),~i=1,2,...,N.
%\end{aligned}
\end{eqnarray}
Consider the following Lyapunov functional:
\begin{eqnarray}
%\begin{aligned}
\fl V=\sum_{i=1}^{N}\mathbf{e}_i^\top(t)\mathbf{e}_i(t)+\sum_{i=1}^N\sum_{j=1}^N(b_{ij}(t)+d_{ij}-c_{ij})^2\nonumber+\sum_{i=1}^N\frac{1}{k_i}(g_i(t)-\bar{g})^2\nonumber\\
\fl+\int_{t-\tau(t)}^t\frac{M}{1-\mu}\sum_{i=1}^{N}\mathbf{e}_i^\top(\theta)\mathbf{e}_i(\theta)d\theta,
%\end{aligned}
\end{eqnarray}
where $\bar{g}$ is a sufficiently large positive constant to be determined.
Thus the diffusion operator $\mathcal{L}$ defined in (\ref{L-op}) onto the function $V$ along with the error system (\ref{d-error}) is:
\begin{eqnarray}
%\begin{aligned}
\fl \mathcal{L}V=2\sum_{i=1}^{N}\mathbf{e}_i^T(t)[\mathbf{f}_i(\mathbf{y}_i(t),\mathbf{y}_i(t-\tau(t)))-\mathbf{f}_i(\mathbf{x}_i(t),\mathbf{x}_i(t-\tau(t)))\nonumber+\sum_{j=1}^N(d_{ij}\Gamma \mathbf{y}_j(t)-c_{ij}\Gamma \mathbf{x}_j(t))\nonumber\\
\fl +\sum_{j=1}^Nb_{ij}(t)\Gamma \mathbf{y}_j(t)-g_i(t)\mathbf{e}_i(t)]
-2\sum_{i=1}^N\sum_{j=1}^N(b_{ij}(t)+d_{ij}-c_{ij})\mathbf{e}_i^\top(t)\Gamma\mathbf{y}_j(t)\nonumber\\
\fl +2\sum_{i=1}^N(g_i(t)-\bar{g})\mathbf{e}_i^\top(t)\mathbf{e}_i(t)
+\frac{M}{1-\mu}\sum_{i=1}^N\mathbf{e}_i^\top(t)\mathbf{e}_i(t)\nonumber\\
\fl-\frac {M(1-\dot \tau(t))}{1-\mu}\sum_{i=1}^N\mathbf{e}_i^\top(t-\tau(t))\mathbf{e}_i(t-\tau(t))+\sum_{i=1}^Ntrace(\sigma_i^\top\sigma_i).
%\end{aligned}
\end{eqnarray}

With the well-known inequality $2\mathbf{x}^\top \mathbf{y}\leq \mathbf{x}^\top \mathbf{x}+\mathbf{y}^\top \mathbf{y}$ and Assumption (H2), one obtains
\begin{eqnarray}
%\begin{aligned}
\fl
  2\mathbf{e}_i^\top (t)[\mathbf{f}_i(t,\mathbf{y}_i(t),\mathbf{y}_i(t-\tau(t)))-\mathbf{f}_i(t,\mathbf{x}_i(t),\mathbf{x}_i(t-\tau(t)))]\nonumber \\\leq \mathbf{e}_i^\top(t)\mathbf{e}_i(t)\nonumber+M[(\mathbf{e}_i^\top(t)\mathbf{e}_i(t)+\mathbf{e}_i(t-\tau(t))^\top \mathbf{e}_i(t-\tau(t)))].
%\end{aligned}
\end{eqnarray}
%\begin{equation}
%\begin{aligned}
%R(S_2)&= p_2\cdot S_2=\sum_{i\in \mathcal{I}^+(p_2)}B_i+\beta B_{l'}\\
%&\leq \sum_{i\in \mathcal{I}^+(p_2)}B_i+B_{l'}\leq \sum_{i\in \mathcal{I}^+(p_1)}B_i \\
%&\leq \sum_{i\in \mathcal{I}^+(p_1)}B_i+\alpha B_l=R(S_1)
%\end{aligned}
%\end{equation}
Let $\mathbf{e}(t)=(\mathbf{e}_1^\top(t),\mathbf{e}_2^\top(t),...,\mathbf{e}_N^\top(t))^\top$, then
\begin{eqnarray}
%\begin{aligned}
\fl \mathcal{L}V\leq(1+M)\sum_{i=1}^N \mathbf{e}_i^\top(t)\mathbf{e}_i(t)+M\sum_{i=1}^N \mathbf{e}_i^\top(t-\tau(t))\mathbf{e}_i(t-\tau(t))+2\sum_{i=1}^{N}\sum_{j=1}^{N}c_{ij}\mathbf{e}_i^\top\Gamma \mathbf{e}_j(t)\nonumber\\
\fl -2\sum_{i=1}^{N}\bar {g}\mathbf{e}_i^\top(t)\mathbf{e}_i(t)+\frac {M}{1-\mu}\sum_{i=1}^N \mathbf{e}_i^\top(t)\mathbf{e}_i(t)-\frac{M(1-\dot \tau(t))}{1-\mu}\sum_{i=1}^{N}\mathbf{e}_i^\top(t-\tau(t))\mathbf{e}_i(t-\tau(t))\nonumber\\
\fl +\sum_{i=1}^Ntrace(\sigma_i^\top\sigma_i)\nonumber\\
\fl =(1+M+\frac{M}{1-\mu})\mathbf{e}^\top(t)\mathbf{e}(t)+M\mathbf{e}^\top(t-\tau(t))\mathbf{e}(t-\tau(t))+2\mathbf{e}^\top(t)\mathbf{P}\mathbf{e}(t)-2\bar g\mathbf{e}^\top(t)\mathbf{e}(t)\nonumber\\
\fl  -\frac{M(1-\dot\tau(t))}{1-\mu}\mathbf{e}^\top(t-\tau(t))\mathbf{e}(t-\tau(t))+p\mathbf{e}^\top(t)\mathbf{e}(t)+q\mathbf{e}^\top(t-\tau(t))\mathbf{e}(t-\tau(t))\nonumber\\
 \fl \leq(1+M+\frac{M}{1-\mu}+2\lambda_{max}(\frac{\mathbf{P}^\top+\mathbf{P}}{2})-2\bar g)\mathbf{e}^\top(t)\mathbf{e}(t)\nonumber\\
 \fl+\frac{M(\dot\tau(t)-\mu)}{1-\mu}\mathbf{e}^\top(t-\tau(t))\mathbf{e}(t-\tau(t))+p\mathbf{e}^\top(t)\mathbf{e}(t)+q\mathbf{e}^\top(t-\tau(t))\mathbf{e}(t-\tau(t)),
% \end{aligned}
\end{eqnarray}
where $\mathbf{P}=\mathbf{C}\otimes \Gamma$.

From Assumption (H3), one has $\frac{\dot\tau(t)-\mu}{1-\mu}\leq 0$, which results in
%\begin{eqnarray}
%\begin{aligned}
%&\mathcal{L}V\leq(1+2\lambda_{max}(\frac{\mathbf{P}^\top+\mathbf{P}}{2})+\frac{2M-\mu M}{1-\mu}\nonumber\\
%&+p-2\bar g)\mathbf{e}^\top(t)\mathbf{e}(t)+q\mathbf{e}^\top(t-\tau(t))\mathbf{e}(t-\tau(t)),
%\end{aligned}
%\end{eqnarray}
%\begin{eqnarray}
%\begin{aligned}
\begin{eqnarray}
%\begin{aligned}
\fl \mathcal{L}V\leq(1+2\lambda_{max}(\frac{\mathbf{P}^\top+\mathbf{P}}{2})+\frac{2M-\mu M}{1-\mu}+p-2\bar g)\mathbf{e}^\top(t)\mathbf{e}(t)\nonumber+q\mathbf{e}^\top(t-\tau(t))\mathbf{e}(t-\tau(t))\nonumber\\
\fl \triangleq-\omega_1(\mathbf{e}(t))+\omega_2(\mathbf{e}(t-\tau(t))).
%\end{aligned}
\end{eqnarray}

Let
\begin{eqnarray}
 \bar g>g^*\triangleq\frac{1}{2}(1+2\lambda_{max}(\frac{\mathbf{P}^\top+\mathbf{P}}{2})+\frac{2M-\mu M}{1-\mu}+p+q),
\end{eqnarray} one gets
$\omega_1(\mathbf{e})>\omega_2(\mathbf{e})$ for any $ \mathbf{e}\neq 0$. Moreover,
$\lim\limits_{\parallel \mathbf{e}\parallel \to +\infty}\inf\limits_{0\leq t\leq \infty}{V=\infty}$.
From Lemma \ref{lem1}, one obtains
$\lim\limits_{t \to \infty }{\mathbf{e}(t;\xi)=0}$ a.s. for any initial data $\xi \in C_\mathcal{F_0}^\mu([-\tau,0],\mathbb{R}^n)$. This means that CS of the duplex network
(\ref{new-driven}) and (\ref{new-response}) can be almost surely achieved for almost every initial data.       This completes the proof.

\begin{rmk}
In the duplex,  the drive layer (\ref{new-driven}) and the response layer (\ref{new-response}) may have  different topologies. In addition,
the configuration matrices $\mathbf{C}$  and $\mathbf{D}$   are not necessarily symmetric or irreducible, which means
that the intra-layer topologies  can be undirected or directed, and they may also contain isolated nodes and disconnected clusters. Therefore, the control scheme can be applied to a wide range of duplex networks with unidirectional couplings.
\end{rmk}

\begin{rmk}\label{remark_3}
It is obvious  that when   CS
between the two layers (\ref{new-driven}) and (\ref{new-response}) is almost surely realized, one has  $\mathbf{e}_i(t)\rightarrow 0$ as $t\rightarrow \infty$ for $i=1,2,...,N$. Furthermore, it renders $\dot g_i(t)\rightarrow 0$ and $\dot b_{ij}(t)\rightarrow 0$   for $i,j =1,2,...,N$.  This means that $g_i(t)$ and $b_{ij}(t)$ will almost surely become  constant.
\end{rmk}

Based on Theorem \ref{thm}, one can easily derive the following corollaries.
%corollary 2
\begin{cor}
Assume that (H1) (H2) and (H3) hold. If the two layers have identical configuration matrices ($\mathbf{C}=\mathbf{D}$), then the drive layer (\ref{new-driven}) and response layer (\ref{new-response}) can  almost surely reach counterpart synchronization through the following simplified adaptive control:
$\mathbf{u}_i(t)=-g_i(t)\mathbf{e}_i(t),\dot g_i(t)=k_i\|\mathbf{e}_i(t)\|.$
\end{cor}
\begin{cor}
Assume that (H1) (H2) and (H3) hold. If there is no noise perturbation,  the duplex network  (\ref{new-driven}) and   (\ref{new-response}) can reach counterpart synchronization through the following adaptive control:
$\mathbf{u}_i(t)=\sum_{j=1}^{N}b_{ij}(t)\Gamma \mathbf{y}_j(t)-\mathbf{g}_i(t)\mathbf{e}_i(t),\dot g_i(t)=k_i\|\mathbf{e}_i(t)\|^2, \dot b_{ij}(t)=-\mathbf{e}_i^\top\Gamma \mathbf{y}_j(t).$
\end{cor}

\section{Numerical simulations}\label{simulations}
In this section, two examples are given to illustrate the feasibility and effectiveness of the proposed  synchronization scheme.
\begin{exam}
Consider a duplex network, each  layer being composed of 5   nodes. The chaotic L\"{u} system with various  parameters is taken as node dynamics, with  the $i-$th ($i=1,2,...,5$) node in both layers being described by
\begin{eqnarray}
%\begin{aligned}
\fl \mathbf{\dot x}_i=\mathbf{f}_i(t,\mathbf{x}_i(t),\mathbf{x}_i(t-\tau(t)))
=
\left(
\begin{array} {ccc}
  (36+i*0.1)(x_{i2}(t)-x_{i1}(t)) \\
  -x_{i1}(t-\tau)x_{i3}(t-\tau)+20x_{i2}(t)\\
  x_{i1}(t-\tau)x_{i2}(t-\tau)-3x_{i3}(t)\\
\end{array}
\right)
  %\end{array}
%\right)
\\
\fl=
\left(
\begin{array} {ccc}
-(36+i*0.1) \quad& 36+i*0.1 \quad& 0\\
0 \quad& 20 \quad& 0 \\
0 \quad& 0 \quad& -3\\
\end{array}
\right)
\left(
\begin{array} {l}
  x_{i1} \\
  x_{i2}  \\
  x_{i3}  \\
\end{array}
\right)
+
\left(
\begin{array} {c}
  0 \\
  -x_{i1}(t-\tau)x_{i3}(t-\tau)  \\
  x_{i1}(t-\tau)x_{i2}(t-\tau)  \\
\end{array}
\right)\nonumber
 \\ \triangleq H\mathbf{x}_i(t)+G(\mathbf{x}_i(t-\tau)).
%\end{align*}
%\end{aligned}
 \iffalse
 Let
%\begin{align*}
~G(\mathbf{y})=
\left(
\begin{array} {c}
  0 \\
  -y_1y_3  \\
  y_1y_2  \\
\end{array}
\right). \fi
\end{eqnarray}
%\end{align*}
%\begin{eqnarray*}
 Since the L\"u system is  chaotic, it is bounded in a certain region \cite{Li2006}.
Thus  there exists a positive constant $R$
such that $\|y_k\|\leq R$ and $\|z_k\|\leq R$ for $k=1,2,3$. Therefore, one has
\begin{eqnarray}
\fl \|G(\mathbf{y})-G(\mathbf{z})\|=\sqrt{[z_3(y_1-z_1)+y_1(y_3-z_3)]^2+[y_1(y_2-z_2)+z_2(y_1-z_1)]^2}\nonumber\\
 \leq \sqrt{2}R\|\mathbf{y}-\mathbf{z}\|.
\end{eqnarray}
That is to say, Assumption (H2) is satisfied with $M=\sqrt{2}R$ for $i=1,2,...,5$.

 The configuration matrices  $\mathbf{C}$ and $\mathbf{D}$ for the drive  layer and response layer are given as
 \begin{eqnarray}
%\left)
%\begin{array} {llll}
 \fl  \mathbf{C}
  %\end{array}
%\right)
=
\left(
\begin{array} {lcccr}
  -6&2&0&3&1\\
  3&-4&1&0&0\\
  0&1&-4&3&0\\
  3&0&3&-7&1\\
  1&0&0&2&-3\\
\end{array}
\right)  \mbox{and}~
 \mathbf{D}
  %\end{array}
%\right)
=
\left(
\begin{array} {lcccr}
  -3&0&2&0&1\\
  0&-6&1&3&2\\
  3&1&-4&0&0\\
  0&1&0&-3&2\\
  1&2&0&1&-4\\
\end{array}
\right),
\end{eqnarray}
%\begin{eqnarray}
%%\left)
%%\begin{array} {llll}
%  \mathbf{D}
%  %\end{array}
%%\right)
%=
%\left(
%\begin{array} {lcccr}
%  -3&0&2&0&1\\
%  0&-6&1&3&2\\
%  3&1&-4&0&0\\
%  0&1&0&-3&2\\
%  1&2&0&1&-4\\
%\end{array}
%\right),
%\end{eqnarray}
respectively. The inner coupling matrix is taken as $\Gamma$=[1 1 0;0 1 0;0 0 1] and node delay  is $\tau=0.003$.
Take $\sigma_i(t,e_i,e_i(t-\tau))=\sigma_0diag(e_{i1}(t)-e_{i1}(t-\tau),e_{i2}(t)-e_{i2}(t-\tau),e_{i3}(t)-e_{i3}(t-\tau)), \sigma_0=1$~ for $i=1,2,...,5$, then $\sigma_i(t,e_i,e_i(t-\tau))$ satisfies the Lipschitz condition and the linear growth condition. That is,  $trace(\sigma_i^T\sigma_i)\leq 2\sigma_0^2e_i^T(t)e_i(t)+2\sigma_0^2e_i^T(t-\tau)e_i(t-\tau)$. Meanwhile, assume that $w(t)=[w_1(t),w_2(t),w_3(t)]$ is a three-dimensional Brownian motion. The initial values of the $i-$th nodes in the drive and response layers are set to be $(x_{i1}(t),x_{i2}(t),x_{i3}(t))=(1+0.3i,-0.6+0.3i,0.3+0.3i)$, and $(y_{i1}(t),y_{i2}(t),y_{i3}(t))=(1-\sin i,1-0.3\cos i,-0.3i),t\in [-\tau,0]~(i=1,2,...,5)$, respectively. The initial values of adaptive gains $g_i(t)~(i=1,2,...,5)$ and adaptive parameters $b_{ij}(t)( i,j=1,2,...,5)$ are chosen randomly in (0,1).

Figure \ref{error1} shows  the counterpart synchronization error of the duplex network (\ref{new-driven}) and (\ref{new-response}). The left panel shows $e_{ij}(t)$, while the right panel shows  the total synchronization error  $\parallel e \parallel=\sqrt{\sum_{i=1}^5\sum_{j=1}^3(y_{ij}(t)-x_{ij}(t))^2}$. It is obvious that CS is almost surely achieved once the proposed control scheme is employed.  Figure \ref{feedback1} further displays the adaptive feedback gains $g_i(t)~(i=1,2,...,5)$ and adaptive parameters $b_{ij}(t)~(i,j=1,2,...,5.)$ varying with time. It is seen that all the parameters reach constant values, which is consistent with  Remark \ref{remark_3}.

%In this section, OS between two-layer scale-free networks are considered. The size of each layer is assumed to be N = 200 and network models are generated by the well-known BA
%algorithm.  Parameters setting are almost the same with above example , the changing is setting dimension  to 200. Figure 3 shows  the total synchronization error  between drive layer  (\ref{new-driven}) and  response layer (\ref{new-response}) on top of scale-free networks.

%\begin{gather*}
%C=
%\begin{pmatrix}
%  -6&2&0&3&1\\
%  3&-4&1&0&0\\
%  0&1&-4&3&0\\
%  3&0&3&-7&1\\
%  1&0&0&2&-3\\
%\end{pmatrix}
%D=
%\begin{pmatrix}
%  -3&0&2&0&1\\
%  0&-6&1&3&2\\
%  3&1&-4&0&0\\
%  0&1&0&-3&2\\
%  1&2&0&1&-4\\
%\end{pmatrix}
%\end{gather*}
%\begin{figure}
%% Use the relevant command to insert your figure file.
%% For example, with the graphicx package use
%\includegraphics[width=0.8\textwidth]{11}
%\caption{Trajectories of the synchronization error between networks (4) and (5)}
%\label{fig:11.fig}       % Give a unique label
%\end{figure}

%\begin{figure}[!h]
%\begin{center}
%\includegraphics [width=4.1cm]{12.jpg}
%\caption{Trajectories of the synchronization error between networks (4) and (5)}
%\end{center}
%\end{figure}

\begin{figure}[!hbt]
 % \begin{tabular}{c}
  \centering
  \mbox{\includegraphics*[width=0.42\columnwidth,clip=0]{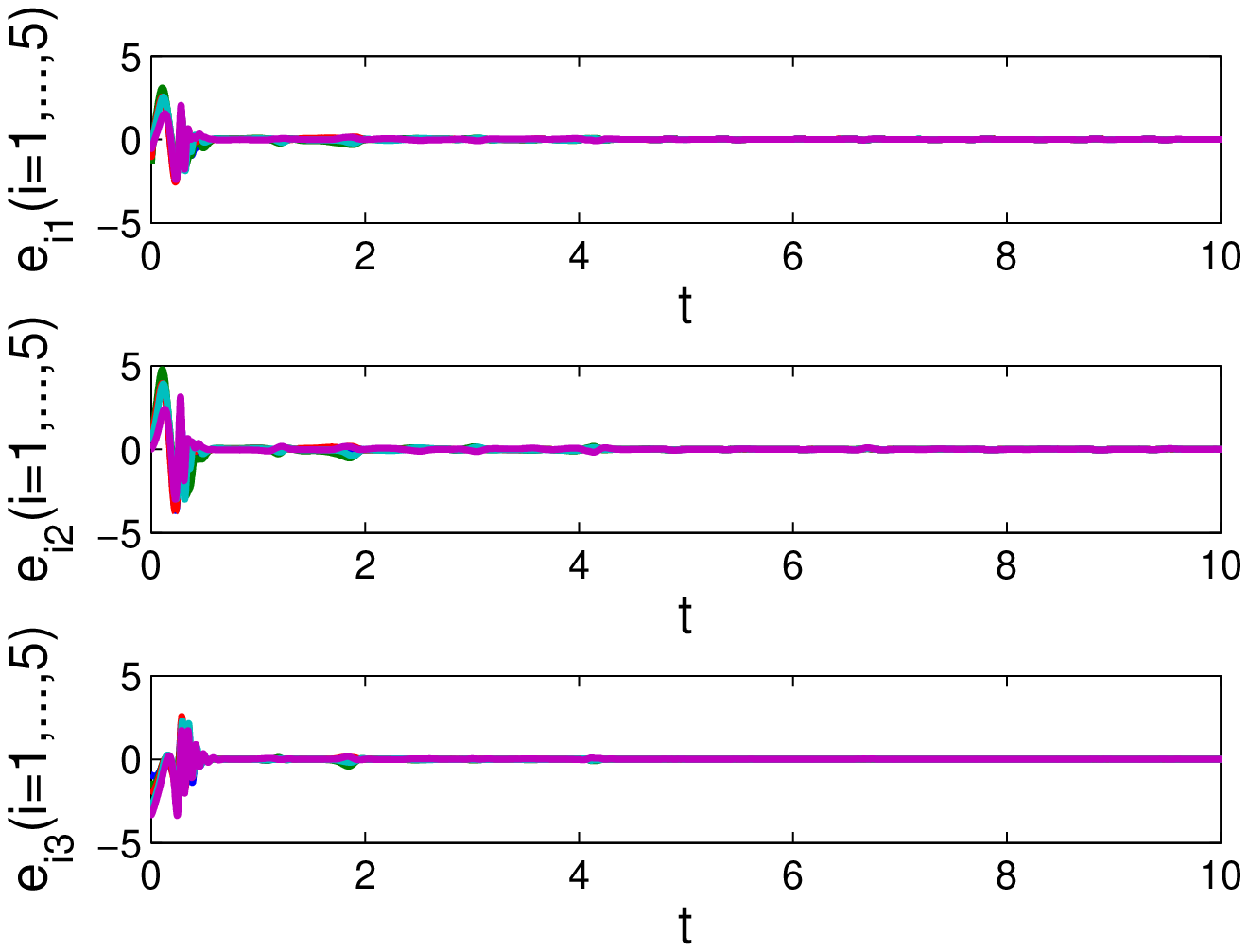}}
  \mbox{\includegraphics*[width=0.42\columnwidth,clip=0]{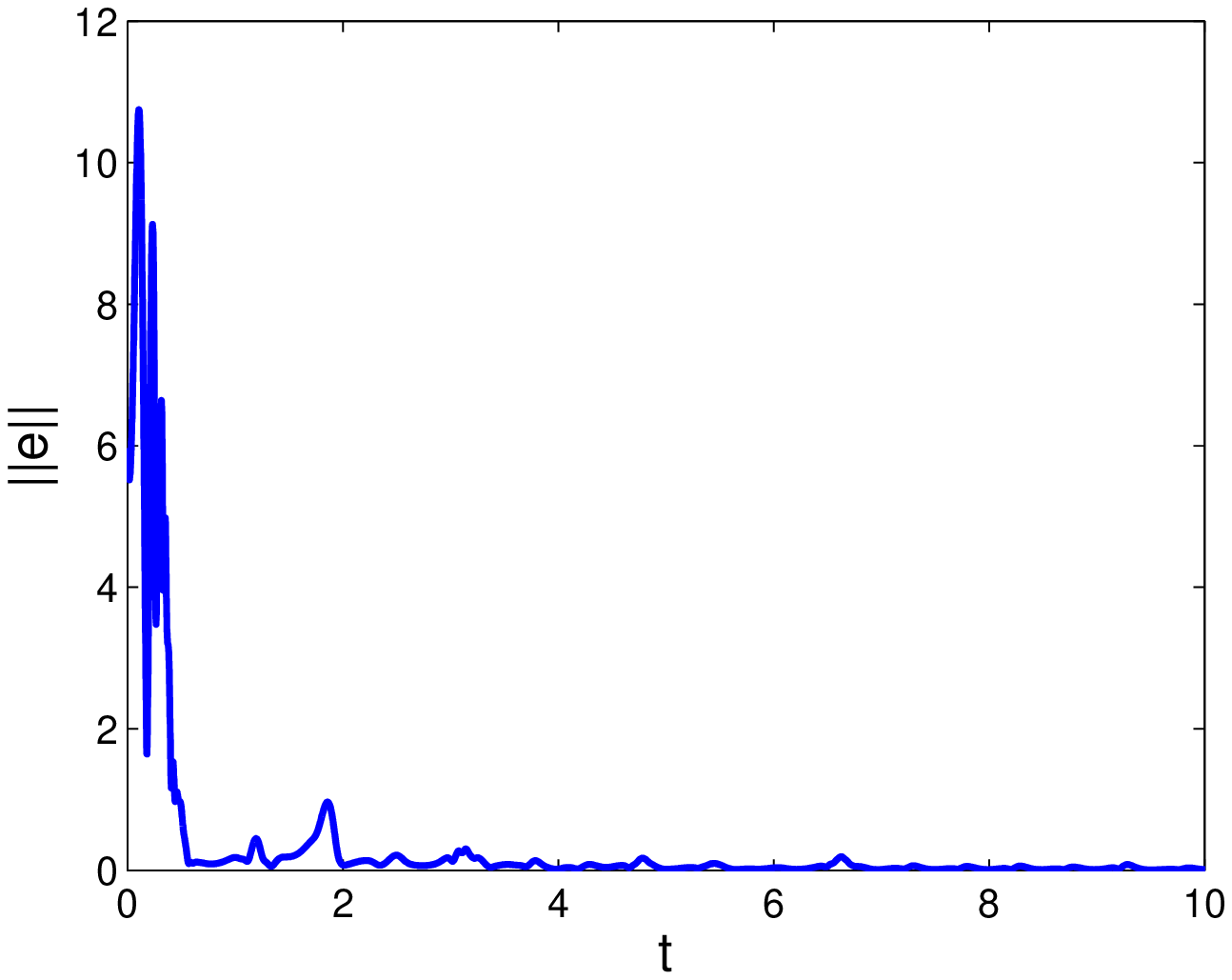}}

   %\mbox{\includegraphics*[width=0.45\columnwidth,clip=0]{11.eps}}
%\end{tabular}
\caption{The counterpart synchronization error between the drive layer (\ref{new-driven}) and the response layer (\ref{new-response}) formed by   L\"u oscillators. Left: $e_{ij}(t)$ varying with time $t$; right:  the total synchronization error.}\label{error1}

  %\includegraphics[width=4.2cm]{11.jpg}\hspace*{3ex}
 % \includegraphics[width=8.4cm]{12.jpg}
%  \caption{Total synchronization error between networks (\ref{new-driven}) and (\ref{new-response}).}\label{E0}
%  \centering
%%\begin{center}
%\includegraphics[width=8.0cm ]{13.jpg}\hspace*{3ex}
%\includegraphics[width=8.0cm ]{14.jpg}
%\includegraphics[ width=6cm]{ouhe_aver.eps}
\end{figure}

\begin{figure}[ht]
%\begin{tabular}{cc}
\centering
  \mbox{\includegraphics*[width=0.40\columnwidth,clip=0]{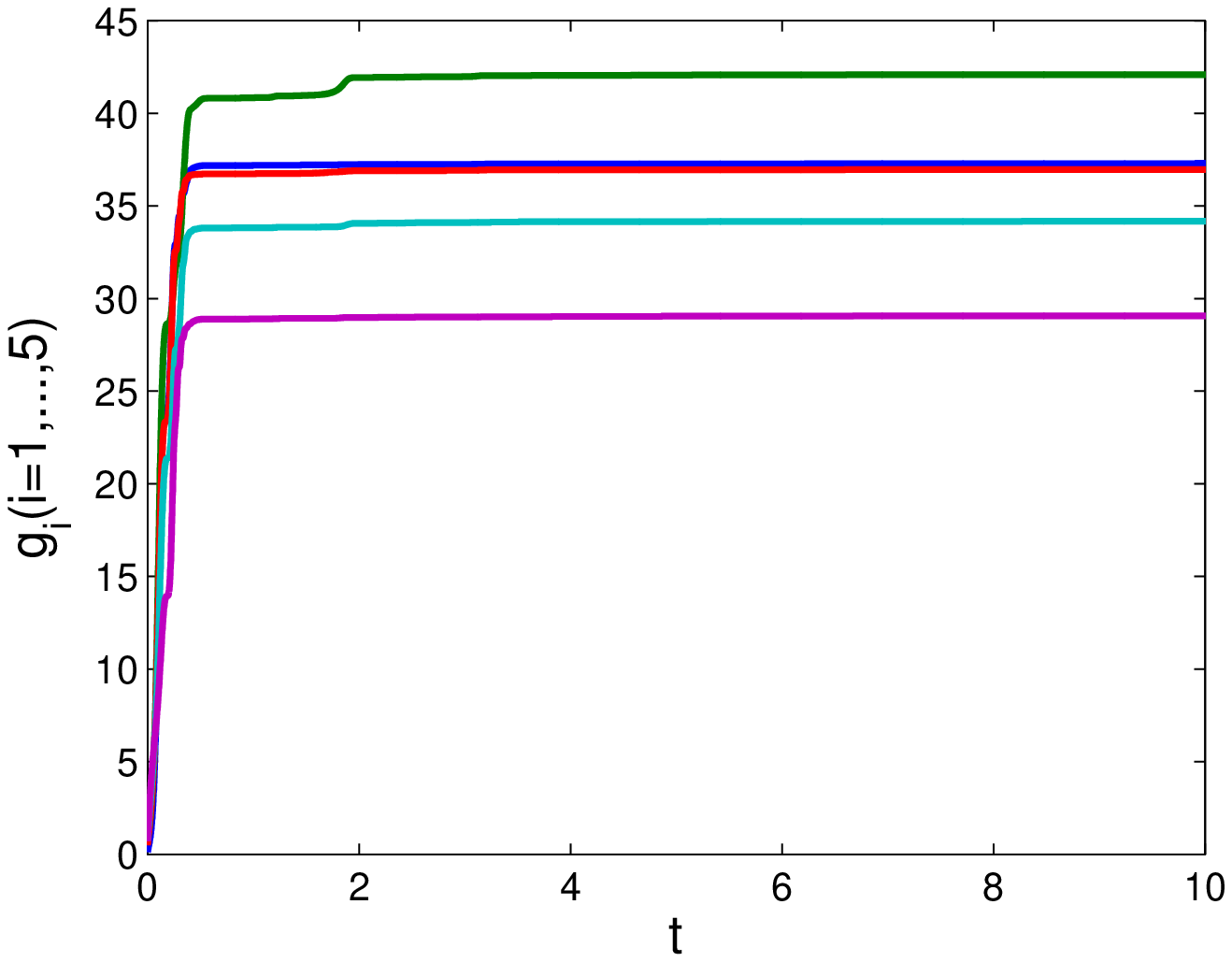}}
  \mbox{\includegraphics*[width=0.40\columnwidth,clip=0]{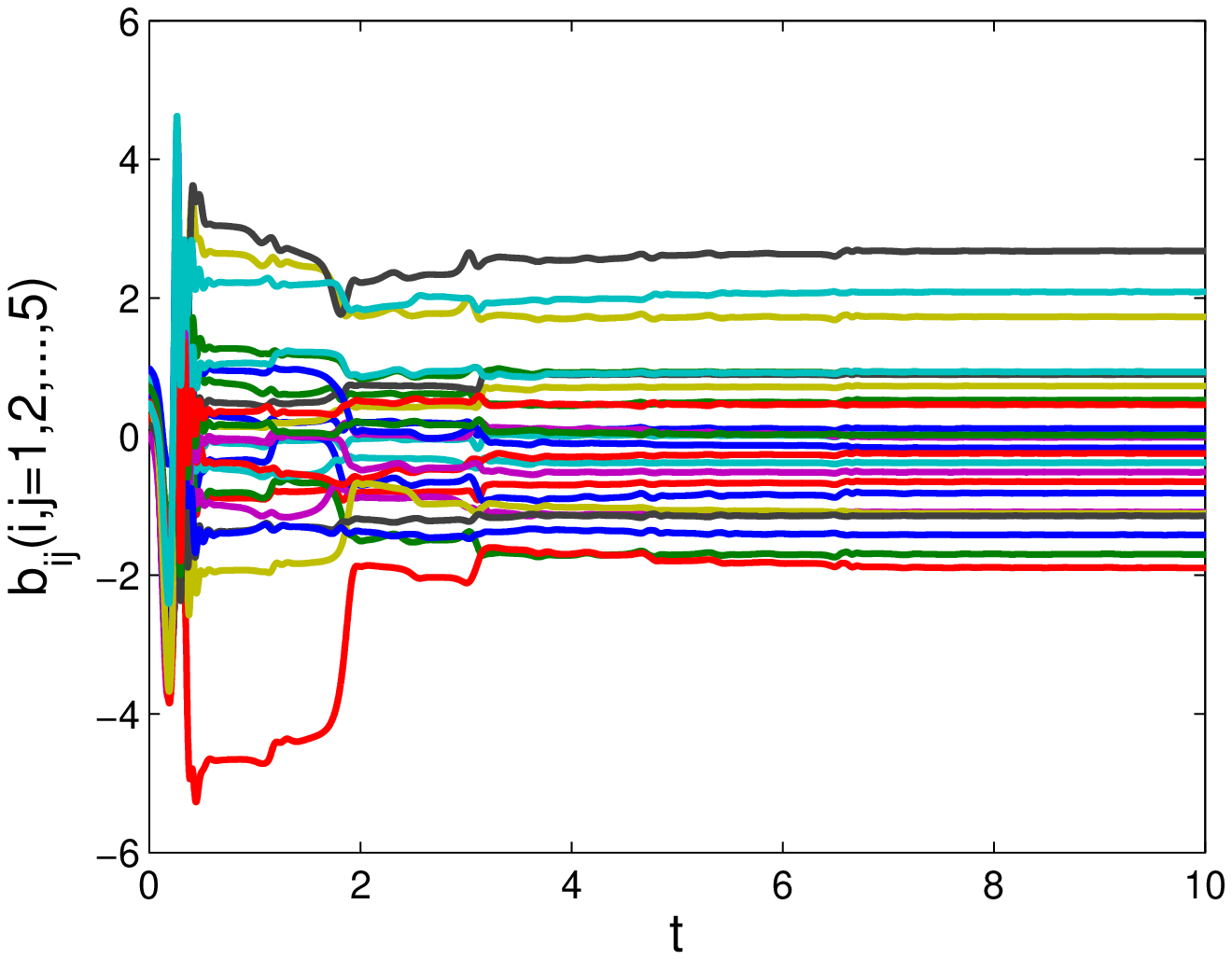}}
%\end{tabular}
\caption{ The adaptive feedback gains $g_i(t)~(i=1,2,...,5)$ (left) and   parameters $b_{ij}(t)~(i,j=1,2,...,5)$  updating according to  (\ref{adaptive})~(right).}\label{feedback1}
\end{figure}

\begin{figure}[ht]
%\begin{tabular}{cc}
\centering
    \mbox{\includegraphics*[width=0.40\columnwidth,clip=0]{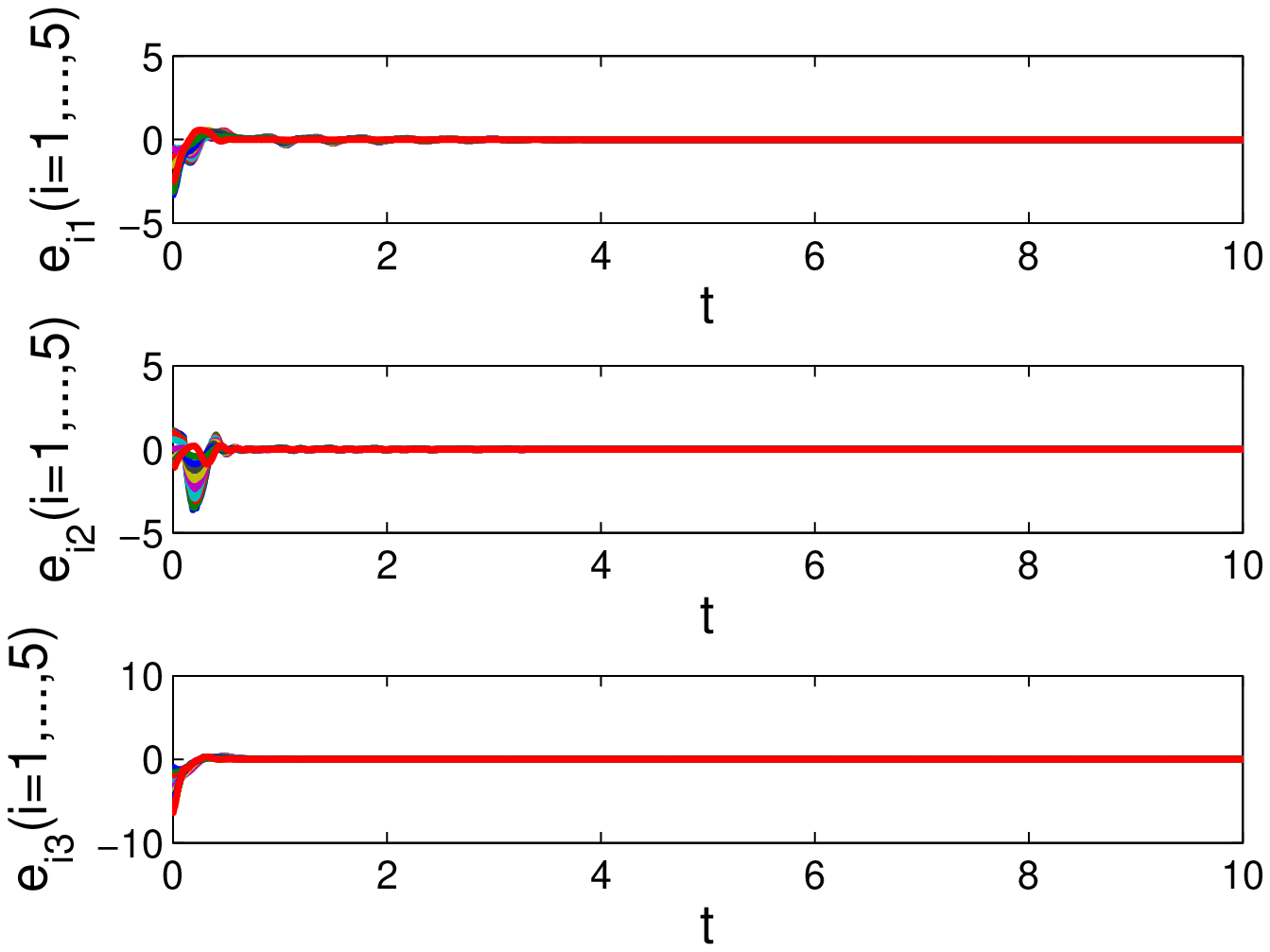}}
  \mbox{\includegraphics*[width=0.40\columnwidth,clip=0]{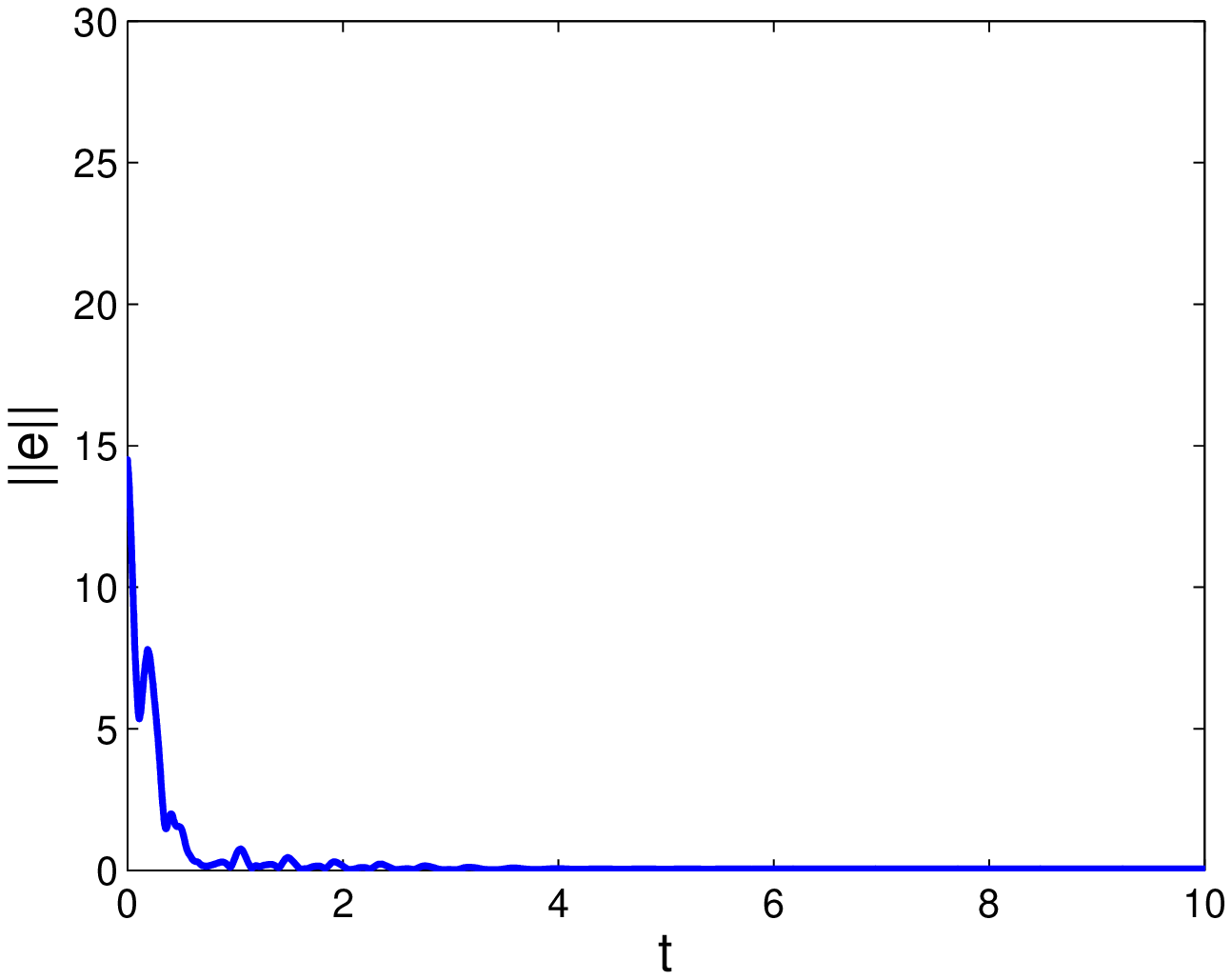}}

 % \mbox{\includegraphics*[width=0.45\columnwidth,clip=0]{23.eps}}
%\end{tabular}
\caption{The counterpart synchronization error of a duplex Hindmarsh-Rose neural network.   Left: $e_{ij}(t)$ varying with time $t$; right:  the total synchronization error.
 }\label{error2}
\end{figure}
\end{exam}
%example2

\begin{exam}
Synchronization   in neuronal networks is one of the burning problems in neuroscience in recent years, and the
Hindmarsh-Rose model \cite{LaRosa2000} has become a popular model for analysis of neuronal activity and has also been extensively
investigated. For example, Fang et al. investigated   chaotic synchronization of nearest-neighbor diffusive
coupling Hindmarsh-Rose neural networks in noisy environments \cite{Fang2009}, and Zhou et al. discussed  the Hindmarsh-Rose model by using impulsive pinning control \cite{Zhou2012}. In what follows, we will discuss the synchronization between two coupled Hindmarsh-Rose neuronal networks. The Hindmarsh-Rose model can be described by a three-dimensional nonlinear differential equations as follows \cite{LaRosa2000}:
\begin{eqnarray*}
%\begin{aligned}
&\mathbf{\dot x}_i=\mathbf{f}(t,\mathbf{x}_i(t),\mathbf{x}_i(t-\tau(t)))\\
&=
\left(
\begin{array}{lll}
  x_{i2}(t)-x_{i3}(t)-x_{i1}(t-\tau)^3+3x_{i1}(t-\tau)^2+I \\
  1-x_{i2}(t)-5x_{i1}(t-\tau)^2  \\
  \mu (4(x_{i1}(t)+\bar x)-x_{i3}(t))  \\
\end{array}
\right).
%\end{aligned}
%\end{eqnarray}
%\end{aligned}
\end{eqnarray*}
Take $I=3,\bar x=1.56, \mu =0.006,\sigma_0=1, \tau=0.1$.   Assumption (H2) is satisfied \cite{Sun2013}. The inner coupling matrix $\Gamma =[1~ 1~ 0;0~ 1 ~0;0~ 0 ~1]$. The intra-layer topologies, the noise term and initial states of   nodes are taken as the same as those in the previous example. Figure \ref{error2} shows    counterpart synchronization errors between two unidirectionally connected  Hindmarsh-Rose networks. Figure \ref{feedback2} further presents the updated feedback gains $g_i(t)~(i=1,2,...,5)$ and adaptive parameters $b_{ij}(t)~(i,j=1,2,...,5)$. It is clearly seen that the numerical simulations perfectly match the theoretical results.

%\begin{figure}[!hbt]
%  \centering
%  % Requires \usepackage{graphicx}
%  %\includegraphics[width=4.2cm]{21.jpg}\hspace*{3ex}
%  %\includegraphics[width=8.0cm]{22.jpg}
%%  \caption{Total synchronization error between networks (\ref{new-driven}) and (\ref{new-response}).}\label{E0}
%%  \centering
%%%\begin{center}
%%\includegraphics[width=8.0cm ]{23.jpg}\hspace*{3ex}
%%\includegraphics[width=8.0cm ]{24.jpg}
%%\includegraphics[ width=6cm]{ouhe_aver.eps}
%\end{figure}
\begin{figure}[ht]
%\begin{tabular}{cc}
  \mbox{\includegraphics*[width=0.45\columnwidth,clip=0]{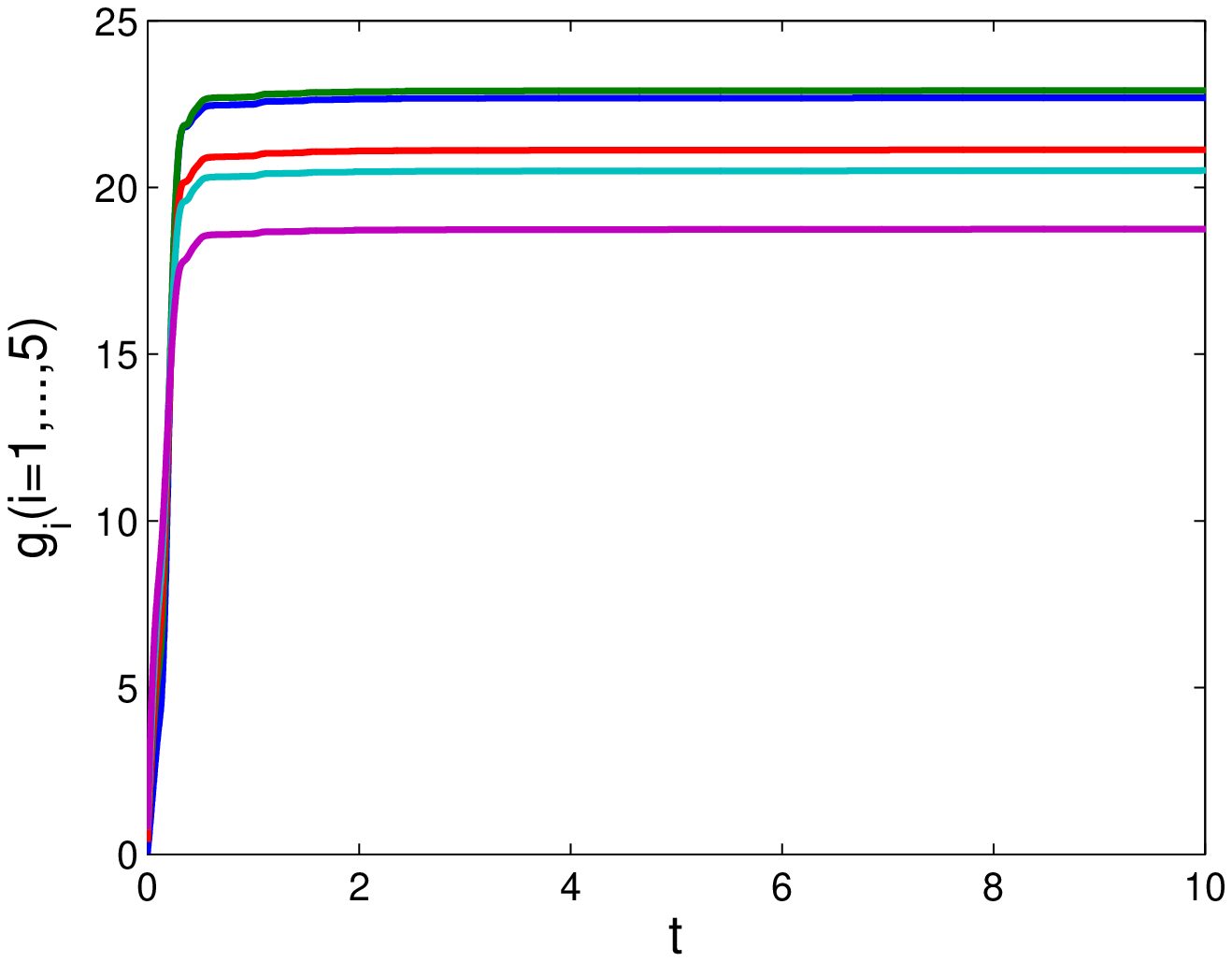}}
  \mbox{\includegraphics*[width=0.45\columnwidth,clip=0]{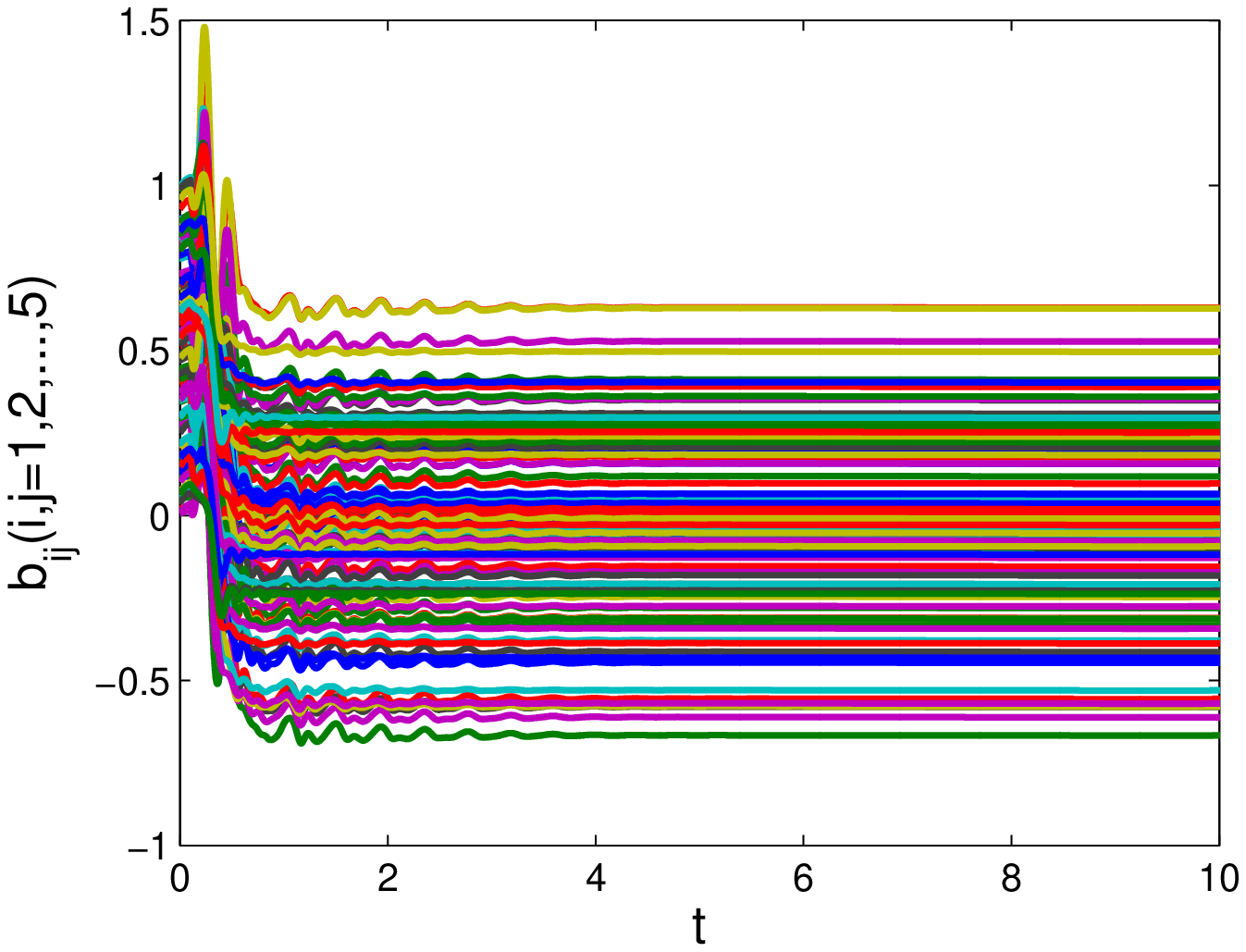}}
%\end{tabular}
\caption{The adaptive feedback gains $g_i(t)~(i=1,2,...,5)$~ (left) and   parameters $b_{ij}(t)~(i,j=1,2,...,5)$ updating according to  (\ref{adaptive})~ (right).}\label{feedback2}
\end{figure}
\end{exam}
\section{Conclusions}\label{con}

In this paper, counterpart synchronization of  duplex networks with delayed nodes and noise perturbation has been investigated. Based on the LaSalle-type invariance principle for stochastic differential equations,  a sufficient condition  guaranteeing CS with the proposed control scheme has been provided.
Numerical examples have also been presented  to illustrate the effectiveness of method. The proposed method will find its applicability to  a wide range of  practical duplex networks.
\section{Acknowledgments}
%The authors would like to thank the editors and the reviewers for their valuable comments and suggestions, which have
%considerably improved the presentation of the paper.
This work is supported by the National Natural Science Foundation of China under Grants 61174028, 61203159, 41201418 and 41301442.
\section{References}
\bibliographystyle{iop}
\bibliography{p2p}

\end{document}